\documentclass[preprint]{aastex}

\newcommand{\Mo}{M_{\odot}} 
\newcommand{\ha}{H$\alpha$\ }

\newcommand{\kms}{km~s$^{-1}$}
\newcommand{\Ho}{$H_0$~= 75~\kms~Mpc$^{-1}$}

\lefthead{Zhang, Q.}
\righthead{ }

\begin{document}

\title{A Multi-Wavelength Study of the Young Star Clusters and Interstellar
       Medium in the Antennae Galaxies}

\author{Qing Zhang\altaffilmark{1,2},
        S. Michael Fall\altaffilmark{1},
        Bradley C. Whitmore\altaffilmark{1}
}

\altaffiltext{1}{Space Telescope Science Institute, 3700 San Martin Drive, 
                 Baltimore, MD 21218, Electronic mail: qzhang@stsci.edu,
                 fall@stsci.edu, whitmore@stsci.edu}
\altaffiltext{2}{Department of Physics and Astronomy, Johns Hopkins 
       University, 3400 N. Charles Street, Baltimore, MD 21218}

\begin{abstract}

We report on a multi-wavelength study of the relationship between young 
star clusters in the Antennae galaxies (NGC~4038/9) and their interstellar
environment, 
with the goal of understanding the formation and feedback effects of 
star clusters in merging galaxies.  This is possible for the first time
because various new observations (from X-rays to radio wavelengths)
have become available in the past several years.  
Quantitative comparisons are made between the positions of the star 
clusters (broken into three age groups) and the properties of the 
interstellar medium by calculating the two-point correlation functions. 
We find that young star clusters are distributed in a clustered fashion, 
demonstrated by power-law angular auto-correlation functions
with slopes in the range $-0.8$ to $-1.0$. 
The young embedded clusters (ages $\sim 5$~Myr) are found to be more 
associated with long-wavelength radiation (mid-infrared and longer), 
while clusters with ages $\sim 10$~Myr or older 
are more associated with short wavelength radiation (e.g., FUV and X-ray).
The youngest star clusters are associated with molecular cloud complexes 
with characteristic radii of about $1$~kpc. In addition,
there is a weak tendency for  them to be found in regions with 
higher HI velocity dispersions.
There is some evidence that both cloud-cloud collisions and shocks 
from recent star formation can trigger star cluster formation, 
but no dominant triggering mechanism is identified for the majority 
of the clusters in the Antennae. 
Feedback from young bright cluster complexes reveals itself in the form of 
large \ha bubbles and \ha velocity gradients in shells around the 
complexes.
We estimate the current star formation rate to be $\approx 20~\Mo$~yr$^{-1}$, 
and the gas consumption timescale to be $\sim 700$~Myr. 
The latter is comparable to the merging time scale and indicates that star 
formation has been enchanced by the merger event.
Finally, we find that the Schmidt law, with index $N\approx -1.4$, 
is also a good description of the cluster formation triggered by merging
in the Antennae. There is some evidence that
feedback effects may modify the Schmidt law at scales below $1$~kpc.

\end{abstract}


%section 1
\section{Introduction}

During the last decade, {\it Hubble Space Telescope (HST)} observations
have revealed the presence of many young star clusters 
formed in various mergers along the Toomre (1977) sequence
(e.g., NGC~4038/9, Whitmore \& Schweizer 1995, 
      Whitmore et al. 1999; 
   NGC~3921, Schweizer et al. 1996; NGC~7252, Miller et al. 1997; 
   NGC~3256, Zepf et al. 1999; 
   see Schweizer 1998 and Whitmore 2001 for reviews), 
as well as a variety of other starburst galaxies (e.g., 
   NGC~1275, Holtzman et al. 1992, Carlson et al. 1998;
   M82, de Grijs, O'Connell, Gallagher 2001).   
These young star clusters are often regarded as candidate
``young globular clusters''.
Indeed, the brightest star clusters formed in mergers have properties (mass, 
size, etc) similar to those of old globular clusters (GCs). 
On the other hand, the luminosity and mass functions of the young star
clusters are completely different from those of old GCs, being power-laws 
for the young star clusters (e.g., Whitmore et al. 1999; Zhang \& Fall 1999,
and references therein), and lognormal or similar centrally-peaked 
distributions for old GCs (with a ``preferred'' scale at 
$M_V\approx -7.3$~mag and $M\approx2\times10^5\Mo$, e.g. Harris 1991).
Theoretical studies show that disruption by various effects might convert an
initial power-law mass function into a centrally-peaked one over a Hubble
time (Vesperini 1998; Baumgardt 1998; Fall \& Zhang 2000).
Thus, the study of young star clusters formed in mergers may increase 
our understanding of the formation of GCs in the early universe.

Irrespective of the GC issue, the discovery of many young star clusters 
in mergers provides a good opportunity to observe cluster evolution 
directly and to test theories that have been proposed for their formation.
For example, clusters might form in the cool and compressed gas behind
strong shocks resulting from the collision of high speed streams 
in the violent environment of merging galaxies (Gunn 1980; Kang et al. 1990).
It has also been suggested that the turbulent velocities between molecular
clouds need to be larger than $50-100$~\kms\ before clusters can form, and higher 
velocity collisions may favor the formation of more massive clusters 
(Kumai, Basu, \& Fujimoto 1993). 
The high speed motions may also lead to high pressure environments that 
trigger turbulence or shocks (Jog \& Solomon 1992; 
Elmegreen \& Efremov 1997; Elmegreen et al. 2000).
Fall \& Rees (1985) have proposed a general two-phase structure due to 
thermal instability and/or shocks that leads to cool clouds compressed 
by surrounding hot gas (see also Kang et al. 1990). 
Finally, star formation may be triggered by stellar 
winds and supernova explosions of young massive stars through compression 
by turbulent motions (e.g., Larson 1993).
We may be able to distinguish between these different models by determining
the correlation between young star clusters and factors in their interstellar 
environment, such as the atomic and 
molecular gas content, the velocity gradient, and the velocity dispersion. 

The Antennae galaxies (NGC~4038/9) provide the nearest opportunity for 
studying young star clusters formed in a prototypical merger. 
A large sample of star clusters with a variety of ages has been
identified using the WFPC2 onboard the {\it HST} (Whitmore et al. 1999).
Following the original discovery of the young clusters, this merger
has been the focus of a full arsenal of state-of-the-art observations 
across a wide range of wavelength bands over the past few years.
These include 
hydrogen 21~cm line emission (Hibbard et al. 2001),
radio continuum at 6~cm and 3.5~cm with the VLA (Neff \& Ulvestad 2000), 
CO~(1-0) observations with the Caltech Millimeter Array 
   (Wilson et al. 2000) and with BIMA (Lo et al. 2000), 
far-infrared emission at 450\micron\ and 850\micron\ with SCUBA 
   (Haas et al. 2000), at 100-160\micron\ (Bushouse, Telesco \& Werner 1998) 
   and at 60\micron\ with the Kuiper Airborne Observatory (KAO) (Evans et al. 1997), 
mid-infrared emission at 15\micron\ with the ISO (Vigroux et al 1996; 
   Mirabel et al. 1998),
and at 10\micron\ with the NASA Infrared Telescope Facility 
   (Bushouse et al. 1998),
\ha line and {\it UBVI} broad-bands with {\it HST} (Whitmore et al. 1999), 
far-ultraviolet at $\sim 1500$~\AA\ with {\it UIT} (Neff et al. 1997), and 
soft X-ray observations with {\it ROSAT} (Fabbiano, Schweizer \& Mackie 1997) and
with {\it Chandra} (Fabbiano, Zezas, \& Murray 2001).
Velocity fields have also been obtained in the hydrogen 21~cm line 
(Hibbard et al. 2001) and \ha line (Amram et al. 1992). 
The wealth of information available on the Antennae provides a 
unique opportunity to conduct a comprehensive study of the relation 
between star clusters and their interstellar environment, and thus provides 
the ideal laboratory to study how star clusters form.
The main limitation in this program is that some observations at longer 
wavelengths have relatively low spatial resolution (e.g., $17\arcsec$
at 60\micron\ and $11.4\arcsec \times 7.4\arcsec$ with hydrogen 21~cm line
emission). 

In this paper, we explore the spatial distribution of star clusters of
different ages and compare these with the intensities observed at other 
wavelengths. 
We quantify these comparisons by calculating two-point correlation
functions. This objective statistic complements, in a quantitative
manner, the subjective impressions one obtains from visually examining
the maps. However, we recognize that this technique does not make full
use of all the available information.
Our goal is to determine the spatial association between star clusters
and the properties of their interstellar environment.  
Some of the questions we hope to explore are: 
What conditions are required for cluster formation~? 
Is there direct evidence showing that high speed gas motion and shocks 
have triggered the formation of young clusters~?  
How is the cluster formation rate related to local gas content~?
Does the Schmidt law hold for star and cluster formation in mergers at 
sub-galactic scales~?

The plan for the remainder of the paper is the following.
In \S2, we divide the sample of clusters into three age groups.
We describe our method for estimating correlation functions in \S3. 
We then calculate in \S4 the autocorrelation functions for star clusters.
In \S~5 and \S6 we cross-correlate the positions of star clusters with  
intensity and velocity maps.
In \S7 we discuss the Schmidt laws for star and cluster formation in the 
Antennae on sub-galactic scales.
Finally, in \S8 we summarize our results and discuss some implications 
relevant to the cluster formation scenarios. 
Throughout, we adopt a distance of $19.2$~Mpc for the Antennae
(\Ho), corresponding to a distance modulus of $31.41$. 

%section 2
\section{Star Cluster Age Groups} 

The star clusters in the Antennae galaxies are found to have a wide spread 
of ages, ranging from $\lesssim 10^6$~years to $\gtrsim 10^{10}$~years 
(Whitmore et al. 1999, and tables therein). In this study, we use the same 
{\it HST} WFPC2 data as in Whitmore et al. (1999). From this cluster sample, 
we form three major age groups of young star clusters: 
  the red clusters (denoted as ``R''), 
  the young bright clusters (denoted as ``B1''), 
  and the older bright clusters (denoted as ``B2'').
The R sample contains candidates for the youngest star clusters, since they 
appear to be still embedded in dusty clouds. The B1 and B2 samples contain
more evolved clusters where much of the surrounding gas has been removed.
We do not consider intermediate-age ($\sim 500$~Myr) or old globular
clusters because they have small sample sizes and they are likely to have 
moved far from where they originally formed.

We define the R clusters as clusters with $V-I>2$. 
To reduce possible stellar contamination
we require the extinction-corrected magnitude to be brighter than $M_V=-10$. 
In estimating the extinction for the R sample, we adopt an intrinsic 
color of $(V-I)_0\approx 0.0$, based on a solar-metallicity cluster less 
than $3$~Myr old in stellar population synthesis models 
(e.g., Leitherer et al. 1999).
Moreover, we exclude individual objects that were identified as foreground 
stars and old globular clusters by Whitmore et al. (1999).

We verify the young ages of the R clusters by comparing them with stellar
population synthesis models in both the $U-B$ vs. $B-V$ diagrams and
the reddening-free $Q$-parameter diagrams, as discussed by Zhang \& Fall (1999).
We first correct the $U$-band magnitudes of the R clusters for the red 
leak of the F336W filter of the WFPC2 by adopting the values 
listed in the WFPC2 manual for M stars with values of $V-I$ similar
to those of the R clusters.
In Figure~1, we show both the stellar population synthesis models 
(Starburst99, Leitherer et al. 1999; Bruzual \& Charlot 1996, hereafter BC96) 
and the objects with errors in both $U-B$ and $B-V$ smaller than 1~magnitude. 
We find that 17 out of the 19 objects (nearly 90\%) are in the regions of the 
diagram indicating young ages (i.e. $Q_1 < -0.7$; ${\rm age}< 10$~Myr).
This confirms the young age of the R clusters. 
The scatter toward more negative values of $Q_1$ may indicate that 
we have not fully corrected for the red leak in the F336W filter.
We exclude the objects with $Q_1>-0.7$ (corresponding to old ages as 
discussed above) from the subsequent analysis.
This results in a sample of $84$ objects, listed in Table~1.

Our R sample includes the well studied cluster \#80 from Whitmore \& Schweizer 
(1995), hereafter WS\#80, which has age estimates of $\sim 4$~Myr (Gilbert
et al. 2000) and $5.5$~Myr (Mengel et al. 2000) from infrared spectroscopy.
We estimate the extinction of WS\#80 to be $A_V=5.7-6.2$ using $V-I$ 
colors and stellar population synthesis models (Leitherer et al. 1999; BC96).  
Other estimates of the extinction are $A_V\sim 9-10$ (Gilbert et al. 2000)
and $A_V=4.3\pm0.3$ (Mengel et al. 2000). 
We note that these young clusters are relatively faint due to 
higher levels of extinction and are thus subject to 
larger observational errors than bright clusters.
This introduces some uncertainty in the separation of heavily obscured
young clusters from red stars, which may not be entirely excluded from 
the R sample.

We identify the B1 and B2 samples with the use of the $Q_1Q_2$ diagram
by the same procedure as in Zhang \& Fall (1999).
By comparing the $Q$ parameters with stellar population synthesis models,
we determine the ages, intrinsic colors, and extinction for the clusters.
We use the BC96 models with a Salpeter IMF and solar metallicity.
We also require the extinction-corrected $M_V$ to be brighter
than $-9$, because almost all known individual stars are fainter than this 
limit (Humphreys 1983).
The mean extinction, $A_V$, varies from $\sim 1.5$ for the youngest B1 clusters
($t\lesssim 10$~Myr) to $\sim 0.3$ for the oldest B2 clusters 
($t\gtrsim 100$~Myr).
The masses of the clusters are mostly between $10^4\Mo$ and $10^6\Mo$ 
(Zhang \& Fall 1999).
The B1 and B2 samples are defined as the clusters with ages 
$3\lesssim t\lesssim 16$~Myr and $16\lesssim t \lesssim 160$~Myr,
and contain 1560 and 327 objects, respectively.
The division at $16$~Myr between the two samples is mainly due to 
a sharp cusp in the tracks of stellar synthesis models in the $Q_1Q_2$-diagram 
(see Figure~1 of Zhang \& Fall 1999). 
The brightest objects in the B1 sample are listed in Whitmore et al. (1999). 
We note that the B1 and B2 clusters are selected according to 
their intrinsic rather than their apparent colors; this procedure minimizes
any selection biases caused by obscuration.

We also identify a sample of candidate young stars for the purpose of comparing
their distribution with those of the clusters (see \S4). 
About $4000$ candidate young stars are selected based on the criteria that
they are relatively faint ($-6.0<M_V < -9.0$) and are located far from 
the stellar population synthesis models in the $Q_1Q_2$-diagram.
Based on their luminosities, most of these objects are probably individual 
massive stars younger than $10^7$~years, although some might be small 
associations where a single star dominates both in brightness and color. 

\section{Correlation Functions}

We will adopt the two-point correlation function to explore the 
distribution of star clusters and their association with flux maps
and velocity fields in other wavelength bands.
This quantitative measure of clustering is useful
as a complement to the visual impression of the maps. 
Generally, the 3D two-point correlation function $\xi(r)$ is defined 
such that $\bar{n}[1+\xi(r)]d^3r$ is the probability of finding a 
neighbor in a shell element with volume $d^3r$ at a distance of $r$ 
from any object in the sample, and $\bar{n}$ is the average density of
objects (cf. Peebles 1980). 
In principle, for the 2D discrete distribution of star clusters, 
the autocorrelation function can be estimated as 
\begin{equation}
1+\xi(r)=\frac{1}{N\bar{n}} \sum_{i=1}^{N} n_i(r)
\end{equation}
where $n_i(r)$ is the number density of objects found in the annulus 
of radius $r$ centered on object $i$, and $N$ is the total number of objects
(e.g., Martinez 1991).
When an annulus extends outside the studied region, a boundary correction is 
applied by only taking into account the included area using a Monte 
Carlo algorithm.
In the case of associating star clusters with a continuous map (e.g., flux or 
velocity gradient maps), the cross-correlation function is estimated, as
\begin{equation} 
1+\xi(r) = \frac{1}{N\bar{f}}\sum_{i=1}^{N} f_i(r). 
\end{equation}
The function $f_i(r)$ is the average flux in an annulus with 
radius $r$ centered on cluster location $x_i$,
and $\bar{f}$ is the average flux over the whole region.

The large-scale structure of the Antennae galaxies introduces a complication 
when calculating the correlation functions for the star clusters. 
Unlike the classical case of determining the correlation function between 
galaxies on the sky (e.g., Peebles 1980), where large-scale uniformity of 
the distribution of galaxies is assumed, the star clusters in the Antennae 
are located in a galaxy with non-uniform large-scale structure.
The simple fact that both the star clusters and 
the emission from the various wavelength bands are only found within
the galaxy introduces a positive correlation, even if there is no other 
physical connection between the star clusters and their interstellar environment.
In the following sections we correct for this effect by subtracting
a smoothed flux map from the original map, both normalized by
their average density.
This is equivalent to the subtraction of $\xi(r)_s$, 
obtained from a smoothed flux map that reflects the large-scale 
structure of the galaxy, from $\xi(r)_o$, obtained from the original map. 
Hence, the ``corrected'' correlation function is given by
\begin{equation}
\xi(r)_c = \xi(r)_o - \xi(r)_s.
\end{equation}
In order to calculate $\xi(r)_s$, we adopt a Gaussian filter with a ${\rm FWHM} 
= 3$~kpc ($\approx 32\arcsec$), which is large enough so that two clusters 
cannot communicate within $100$~Myr at a signal speed of 
$\lesssim 30$~\kms.  
This smoothing scale is much larger than the resolution of the observations 
we use in this study, but is less than a quarter of the scale of the 
WFPC2 field of view. 

To subtract the effects of large-scale galactic structure when calculating the
autocorrelation functions for clusters and candidate young stars,
we use equation~(2) along with the following technique to estimate $\xi(r)_o$
and $\xi(r)_s$.
We first map the individual objects onto a $700\times 700$ grid, which is
the same size as the flux maps. 
The autocorrelation is then calculated between the positions of the 
individual objects and the density distribution defined by the grid map.
The central object on the grid is always excluded to avoid redundant counting 
of the objects.
Finally, since objects are only located within the field of view of the WFPC2, 
we mask out other regions of the map when calculating the autocorrelation 
functions. 

We also cross-correlate the cluster positions with maps of velocity 
gradients and velocity dispersions.
Note that the correlation is not made with the velocity itself, which
would be meaningless.
The velocity gradient at each pixel is calculated by finding the largest 
spread in the integrated line-of-sight velocities  within a rectangular 
box with the equivalent size and orientation of the observed spatial 
resolution beam.
Regions without velocity information are not considered in the 
calculation of the velocity gradient, and hence are not included in 
estimating the correlation functions.
In this case, we do not need to subtract the smoothed correlation
function, as in equation~(3), because the method of deriving 
gradients already removes the large-scale structure of the galaxy.

Finally, we estimate the statistical uncertainties in the correlation 
functions. 
For the auto-correlation functions, we adopt an estimate of uncertainty
$N_p^{-1/2}$, where $N_p$ is the number of distinct pairs
of objects (cf. Peebles 1980, \S48). Although this estimate is strictly
valid only for small $\xi(r)$, it may provide an approximate
indication of the uncertainties when $\xi(r)$ is large.
For the cross-correlation functions we consider only the statistical
uncertainty due to the finite number of clusters, $N$. 
This leads to approximate fractional errors of $1/\sqrt{N}$ in $1+\xi(r)$.
Any errors in the determination of the flux and velocity gradients would
increase the uncertainty. 

With the autocorrelation and cross-correlation functions computed as above, 
a random distribution of clusters will result in a flat correlation, with 
$\xi(r)_c= 0$. On the other hand, a peaked $\xi(r)_c$ at small radii 
indicates a positive correlation. The width of the central peak represents 
the spatial scale of association between the clusters and the flux for the 
various wavelength bands (convolved with the resolution), while the absolute
value of $\xi(r)_c$ is a measure of the concentration of flux surrounding 
the clusters at a given distance $r$, relative to the average over 
the whole galaxy.
From a practical standpoint, it should be noted that the resolution can 
sometimes dominate the determination of the width and amplitude of the peak.

%section 4
\section{The Spatial Distribution of Star Clusters}

We first compare the spatial distributions of the young star clusters 
and the candidate star samples in Figure~2.  
We label several interesting regions, major bright knots and 
the nuclei of NGC~4038 and NGC~4039, for later references.
The R clusters mainly reside in the ``overlap region'' 
(where the main bodies of NGC~4038 and NGC~4039 are superimposed), 
as well as the western loop of NGC~4038.
The B1 and B2 clusters spread more widely over the Antennae galaxies.
We find that they have generally similar spatial distributions, although
the B1 sample is slightly more clustered than the B2 sample.
The distributions of candidate young stars and young clusters are broadly 
similar, but at small scales, they are distinctly different.

The autocorrelation functions for the cluster samples and 
the candidate young stars are shown in Figure~3. 
The panels from the top to the bottom show $\xi(r)_o$, $\xi(r)_s$, and 
$\xi(r)_c$, respectively, as defined by Equation~(2).  
The observed correlation functions decrease approximately as
power laws up to a distance of $\approx 15\arcsec$, corresponding to 
$\approx 1.4$~kpc. Beyond this scale, they decrease even more rapidly.
The smoothed correlation functions, shown in the middle panel, are 
essentially flat within $\approx 10\arcsec$. 
After the subtraction, the corrected correlation functions shown in the
bottom panel remain power-laws within $\approx 8\arcsec$, corresponding to 
$0.74$~kpc.
The power-laws have indices of $-0.83$, $-1.06$, and $-0.89$ for the R, 
B1, and B2 clusters, and are similar to the value $-0.7$ for galaxies 
(Peebles 1980), although the reasons for this are presumably very different.
It is interesting to note that the autocorrelation functions for the 
candidate young stars are flatter (with a power-law index of $-0.41$), 
implying a weaker correlation. 
The relatively small sample size for the R clusters is
responsible for the noisy behavior in the autocorrelation function;
however, it is clear that $\xi(r)$ for the R clusters largely follows 
those of the B1 and B2 clusters (especially at small $r$) rather than 
that of the candidate young stars. 

We also checked the distribution of the intermediate-age clusters 
($t\sim 500$~Myr) and old globular clusters identified by
Whitmore et al. (1999). 
The intermediate-age population is more spread out than the young clusters,
with most of the clusters situated along streams apparently associated 
with the tidal tails and the northwestern loop of NGC~4038. 
The old GCs are mostly found in the disk of NGC~4039.
Neither of these groups show evidence of much clustering, suggesting an effect
of dispersal and mixing after several galactic orbits.

The two-point autocorrelation function provides some insight into 
the physical processes of cluster formation. 
The clustered distribution of young star clusters probably reflects the 
self-similar structure of the interstellar medium within the
molecular cloud complexes. 
The radius of $\approx 1$~kpc, beyond which the correlation functions decline,
is comparable to the size of giant molecular 
cloud complexes in the Antennae ($0.8-1.8$~kpc, Wilson et al. 2000).
Since the decline is also seen before the effects of large-scale structure
are subtracted, it cannot be an artifact caused by the smoothing. 
We also note that the bright cluster complexes, such as knots~G, M, R, S, T 
in the northwest and B, C, D, E, and F in the south of the Antennae,
as marked in Figure~$2$, all have sizes of $\sim 0.5$~kpc 
(see enlarged images in Figures~6 and 7 of Whitmore et al. 1999), 
in agreement with the scale set by the autocorrelation functions, 
especially if we take into account the possible contraction of the cloud
complexes during cluster formation. 

\section{Cross-correlations between Star Clusters and their Environment}

In this section, we explore the cross-correlation between the star 
clusters and the flux from the Antennae galaxies in various wavelength bands.
By studying these correlations we hope to obtain a comprehensive view
of how the physical distribution of the clusters is determined by 
their interstellar environment and to study how this depends upon the
stages of their evolution. In the following, we present the maps in 
order of decreasing wavelengths. 

We first overlay the positions of the three age groups of clusters 
(the R, B1, and B2 samples) on the contours of various flux maps. 
The maps presented here have an adopted pixel scale of $0.1992\arcsec$.
If not mentioned otherwise, the contours are plotted on a  
logarithmic scale. In addition to the direct comparisons from the maps, 
we quantify the spatial association by calculating the cross-correlation 
functions between the clusters and the flux in various wavelength bands. 
We only show the $\xi(r)_c$'s, with the effect of large-scale galactic 
structure removed. 

\subsection{HI Line Emission (21~cm)}

The hydrogen 21~cm line emission provides a direct measurement 
of the properties of atomic hydrogen gas, both its content and kinematics.
The most recent high resolution and high sensitivity 21~cm observations 
of the Antennae galaxies were obtained with the VLA in C+D array 
configuration, as reported by Hibbard et al. (2001). 
The data with the resolution of $11.4\arcsec \times 7.4\arcsec$ beam 
(with the robust parameter set to $-1$)
were kindly made available by J. Hibbard prior to publication. 
They determine that the total atomic gas content is more than
$4.7\times 10^9~\Mo$, and that about $68\%$ of it is in the tidal tails.
Within the central disks, most of the HI gas is located in NGC~4038 and the
overlap region.
Almost all the star and cluster formation occurs in the disks
rather than in the tails (cf. Hibbard et al. 2001; Knierman et al. 2001),
which implies that the presence of atomic hydrogen alone is not 
sufficient to cause star formation. 

Figure 4 shows the cross-correlation between the integrated flux of 
21~cm line emission and the three age groups of young star clusters. 
The clusters and HI gas have broadly similar distributions, 
although most of the clusters lie away from the peaks of HI flux.
On the other hand, most of the prominent star formation regions have a 
nearby reservoir of HI, with an offset ranging from 
$0.2$ to $0.9$~kpc. The positive cross-correlation functions in panel~$d$ 
indicate that cluster formation is related to atomic gas.
The correlation is the strongest for the R clusters and becomes weaker 
as the clusters age.
Note that for the B1 and B2 samples, $\xi(r)$ is essentially flat at scales 
within $10\arcsec$, reflecting the fact that the clusters generally
reside off the peaks of atomic gas concentrations. 

\subsection{Radio Continuum Emission (6 cm)}

Radio continuum emission can come from either synchrotron radiation from
young supernova remnants or thermal radiation from ionized gas in HII regions.
An earlier radio continuum observation of the Antennae detected both diffuse 
emission and thirteen discrete sources (Hummel \& van der Hulst 1986).
More recent high-resolution radio continuum maps (at 6 cm and 4 cm)
were obtained with the VLA in its BnA, CnB, and B configurations
and are reported by Neff \& Ulvestad (2000). 
They identified 115 compact radio sources, one third of which 
are associated with HII regions and the others with supernova remnants.
In the following comparisons we adopt the 6~cm radio continuum map 
that was kindly provided by S. Neff and J. Ulvestad before publication.
The map has been cleaned using natural weighting with a full-resolution of 
$1.72\arcsec \times 1.52\arcsec$ and high sensitivity (RMS is $10.7~\mu$Jy).

We show in Figure 5 the cross-correlations between radio continuum flux 
and the three age groups of star clusters. 
The continuum flux occurs mainly in the overlap region of the galaxies,
with several prominent point-like sources. 
There is also a significant amount of continuum flux from  the two nuclei 
and the western loop of NGC~4038. 
Several of the R clusters are coincident with the peaks of the radio 
continuum flux, especially in the overlap region and part of the western 
loop of NGC~4038. 
Many B1 clusters are located in regions of high continuum flux in the two 
galactic nuclei and the western loop.
However, some B1 clusters in the northern and eastern star formation regions
are not obviously associated with the continuum flux.
Only a small fraction of the B2 clusters coincide with the continuum flux, 
since they are more evolved and their massive stars have already 
gone supernovae.  
The different association of star clusters with the radio flux is clearly 
shown in Figure~5$d$.  
The cross-correlation is strongest for the R clusters
and weakest for the B2 clusters.

We also compare, in Figure 6, the locations of the R clusters 
with the compact radio sources identified by Neff \& Ulvestad (2000)
with flux larger than twice the detection limit.
Also shown are the 200 brightest B1 clusters, with extinction-corrected $M_V$ 
ranging between $-15.7$ and $-11.5$.
Both the R clusters and the compact radio sources have significant numbers
of objects in the overlap region. 
However, less than half of the compact radio sources 
in the overlap region are directly associated with the R clusters.
Some are associated with bright clumps of B1 clusters, such as knots 
B, C, D, E, and F.
As suggested by Neff \& Ulvestad (2000), some of the radio sources that are 
not obviously associated with star clusters in our sample
may be too deeply embedded in dust to be observed in the optical bands.

\subsection{CO (1-0) Line Emission (2.6~mm)}

CO line emission provides a probe of molecular gas within galaxies
since it is related to H$_2$ column density (Strong et al. 1988; Wilson 1995).
In an early study of the Antennae galaxies, Stanford et al. (1990) identified 
three major concentrations of molecular gas, corresponding to the nuclei 
of NGC~4038 and NGC~4039 and the overlap region, with a total flux of 
430~Jy~\kms. Recent high resolution observations have detected a more 
extended distribution of CO in the whole overlap region and in the western 
loop of NGC~4038 (Lo et al. 2000; Wilson et al. 2000; Gao et al. 2001). 
With the Caltech Millimeter Array, Wilson et al. (2000) reported  
910~Jy~\kms\ of CO flux, corresponding to $5.3\times 10^9~\Mo$, and
Gao et al. (2001) detected a total mass of $1.5\times 10^{10}~\Mo$,
both using the CO (1-0) flux to H$_2$ conversion relation $N({\rm H_2})/I_{CO} 
=3.0 \times 10^{20} {\rm cm^{-2}(K~km~s^{-1})^{-1}}$ (Wilson 1995).
Note that molecular gas is mainly found in the disks, where the mass of 
H$_2$ is $\sim 10$ times the mass of HI.
Five supergiant molecular clouds are identified in the overlap region, 
with masses in the range $3-6\times 10^8~\Mo$ and diameters $0.8-1.8$~kpc 
(Wilson et al. 2000). 

In the following, we mainly use the BIMA observations, which
cover the disks of the Antennae galaxies. 
The map we use is reconstructed from CO contours with known levels
(cf. Lo et al. 2000; Plate~6 of Sanders, Surace \& Ishida 1999). 
The spatial resolution is $6.61\arcsec \times 7.77\arcsec$. 
We repeated our analysis using the Wilson et al. (2000) CO
intensity map and found results similar to those reported 
below (e.g., the relative ordering of the correlations for the
three age groups is the same for $r<10''$). We prefer the BIMA map
since it covers a larger fraction of the Antennae galaxies.

Figure 7 shows the spatial distribution of star clusters 
(the R, B1, and B2 samples) overlaid on CO (1-0) flux contours.
The R clusters show a good correlation with molecular clouds.
The overlap region is where most of the molecular gas is detected,
and where more than half of the R clusters are located. 
Many R clusters are clearly projected on the peaks of CO intensity.
This is also true in the northeastern arm of NGC~4038. 
The R clusters that are not obviously associated with CO (1-0) emission
are located mainly to the north of the NGC~4038 nucleus. 
It is not clear if these objects are genuine embedded clusters,
due to the possible contamination with stars in the R sample. 
We note that it is rare to find R clusters near the concentrations of
molecular gas in the nuclei. 
Most B1 clusters, on the other hand, are located near the peaks of CO flux, 
except for the young clusters near the nucleus of NGC~4038 
(e.g., knots~J and K) and knot~F on the northeastern side. 
The B2 clusters, most of which are on the  northeastern edge or the 
western loop of NGC~4038, are associated with relatively weak CO flux.

The cross-correlation functions between the star clusters and 
molecular gas confirms the above visual impression (Figure~7$d$). 
The strongest correlation is for the youngest clusters (the R sample)
with a half-maximum of $\xi(r)$ at $r\approx 13\arcsec$ ($1.2$~kpc), 
consistent with the size of molecular clouds measured by Wilson et al. (2000). 
The coincidence of R clusters with CO flux supports the idea that
molecular gas is directly associated with star cluster formation.
The correlation is the weakest for the B2 clusters, with a plateau 
in $\xi(r)$ at $r\approx 5.5\arcsec$. One possibility for this plateau 
is that the molecular gas is being consumed, blown away, or dissociated upon 
the formation of star clusters (e.g., the eastern star formation region, 
see \S5.9). 
It is also possible that the molecular clouds migrate differently from 
the clusters within the galaxy. 
A velocity difference of $\sim 5$~\kms\ would result in a separation of 
$\sim~500$~pc (corresponding to $5.4\arcsec$) within $100$~Myr. 
Hence, this plateau in $\xi(r)$ for the B2 sample can possibly result 
from the drifting of the clusters.

\subsection{Far-Infrared Emission ($60\micron$)}

Far-infrared emission is thought to originate mainly from dust grains mildly 
heated either by UV photons from massive stars or by interstellar shocks. 
Emission at 60\micron\ is found to correlate tightly with the radio continuum 
for a wide variety of galaxies (Knapp 1990), suggesting massive star formation 
as the dominant underlying physical cause (Binney \& Merrifield 1998). 
Thus, far-infrared observations should be a valuable tool for revealing the 
physics of star cluster formation in the Antennae, 
a prominent starburst containing a large amount of dust. 

The Antennae galaxies were observed at 60\micron\ with the 90~cm Kuiper 
Airborne Observatory (KAO) with a resolution of $17\arcsec$, corresponding
to $1.58$~kpc, as reported by Evans, Harper, \& Helou (1997). Half of the 
total detected flux, 48.7~Jy, is from the overlap region and a quarter from 
the NGC~4038 nucleus. This is a situation similar to the radio continuum 
emission, CO~(1-0) line emission, and mid-infrared emission (see \S5.5).
The Antennae were also observed at longer wavelengths (SCUBA 450\micron\ and
850\micron), and both warm ($\approx 30$~K, typical for starbursts) and cool 
dust ($\lesssim20$~K, typical for quiescent galaxies) were found in the overlap 
region and in the nucleus of NGC~4038 (Haas et al. 2000). 
In the following we use the KAO 60\micron\ image, which was kindly made 
available by R. Evans. 

Figure 8 shows the spatial distribution of the R, B1 and B2 samples 
overlaid on the contours of the 60\micron\ flux.
The R clusters show a good correlation with the $60\micron$ flux, 
while the B1 and B2 clusters match the far-infrared flux in many locations,
except for the eastern star formation region.
This is a location with weak \ha emission (see \S5.7) and both atomic and 
molecular gas almost depleted, as seen in Figures~4 and 7.
The cross-correlation functions in Figure~$8d$ clearly show the positive
association between the age of the star clusters and far-infrared
emission. The strongest correlation is found for the R clusters. This
is consistent with our interpretation that most of the R clusters are
young embedded objects (see Figure 1 and the discussion in \S2),
where emission from dust would be expected. While this correlation may
partly be due to a selection effect, since dust is responsible for both
the red colors of the clusters and the $60\micron$ emission, we note that
the B1 and B2 clusters were chosen on the basis of their reddening-free
Q parameters (i.e., they can be red or blue) and, hence, do not suffer
from this selection effect. In addition, the fact that the B1 sample
shows stronger correlation than the B2 sample (by $2\sigma$) provides
further support for an age dependence for the dust emission.

\subsection{Mid-Infrared Emission ($15\micron$)}

Compared with far-infrared emission, mid-infrared emission is thought to
be associated with warmer dust grains heated by newly formed massive stars. 
Its close association with very young stars makes it an excellent probe for
formation activity, especially when dust obscuration is significant 
and objects are invisible in optical bands.

Observations of the Antennae in the mid-infrared ($12-17\micron$) were
obtained with ISOCAM on the Infrared Space Observatory (ISO) with a 
resolution of $4.5\arcsec$, as reported by Vigroux et al. (1996) and 
Mirabel et al. (1998). Half of the detected mid-infrared flux is from 
the dusty overlap region while the other half is from the two galactic 
nuclei and the western loop of NGC~4038. 
There is little $15\micron$ flux in the northeastern extension.
Overall, the distribution of mid-infrared flux in the Antennae largely 
resembles the radio continuum, CO, and far-infrared maps.
In the following, we correlate the positions of the star clusters with the ISO 
mid-infrared digital image kindly provided by I. F. Mirabel and his colleagues.

Figure 9 shows the spatial distribution of the R, B1, and B2 samples 
overlaid on the contours of mid-infrared flux.
The R clusters are largely coincident with the peaks of mid-infrared flux 
in the overlap region. A good example is WS\#80, which coincides with 
the most luminous infrared source on the ISO map.
The B1 clusters are also well correlated with mid-infrared emission.
The exceptions are the clumps of clusters located in the eastern star 
formation region (discussed in \S5.4) and south of the nucleus of NGC~4039.
The correlation for the B2 clusters is weaker, in that relatively few
objects reside in the overlap region where mid-infrared flux is strong. 

Figure~9$d$ shows the cross-correlation between the clusters and 
mid-infrared flux. 
Evidently, the R and B1 clusters are closely associated with
the $15\micron$ emission. In particular, we note that for the first time 
the B1 sample shows a stronger peak than the R sample.
The relatively strong correlation with mid-infrared emission for young 
clusters (the R and B1 samples) over the relatively weak 
correlation for older ones (the B2 sample)
indicates that dust grains are either heated or formed along with 
massive star clusters, and either cool or are expelled as the clusters age.
The selection effect discussed in \S5.4 is again relevant for $15\micron$
emission, but here we are on firmer ground since the difference between
the B1 and B2 samples are quite dramatic.

\subsection{I-band Emission (0.8\micron)}

So far we have presented the correlations between clusters and long-wavelength
radiation related to their gas and dust environment. 
We now compare the clusters with the distribution of old stars in the host 
galaxies. 
These populations are best depicted by the near-infrared bands because of the
existence of red giants and the absence of hot blue stars.

The Antennae have been observed in both the J ($1.25\micron$) band 
(Bushouse \& Werner 1990) and K ($2.2\micron$) band (Evans et al. 1997).
These images clearly reveal the prominent nuclei and disks of NGC~4038 and 
NGC~4039.  The disks are largely symmetric, except for some bright HII 
regions adorning the overlap region between the two nuclei and the 
western loop of NGC~4038, and an extension of the NGC~4038 disk in the east.
In the following, we use the I-band image of the Antennae obtained with 
the F814W broad-band filter of the WFPC2 on {\it HST} 
(Whitmore et al. 1999). This image has a resolution of $0.0455\arcsec$ in the 
PC chip and $0.0996\arcsec$ in the WF chips (corresponding to $4.23$~pc 
and $9.26$~pc, respectively), and generally resembles
the J and K band images.

We show in Figure 10 the star clusters overlaid on the contours of the 
I-band flux. 
The R clusters do not coincide well with the disk or nucleus of either 
NGC~4038 or NGC~4039, where most of the near-infrared flux is detected. 
The B1 clusters, on the other hand,  spread over the arms of NGC~4038 and 
NGC~4039 as well as the overlap region. 
Density enhancements of the B1 clusters coincide with nearly all the I-band 
point-like sources in the overlap region, the western loop of NGC~4038, 
and even the extended arm structure of NGC~4039 toward the southeast. 
The B1 clusters are also found near both nuclei. 
The situation for the B2 clusters is similar to that of the B1 clusters.
As Figure~$10d$ shows, the R clusters and I-band flux have little 
correlation, contrary to that for the B1 and B2 clusters. 

It is possible that obscuration by dust in the overlap region is 
responsible for the weak correlation between 
the R clusters and the I-band flux.  This is not likely to be
the whole story, however, since the J and K images, with much less
extinction, look very similar to the I-band image. 
The second possibility is that the older clusters formed primarily
in the disks of NGC~4038 and NGC~4039, alongside existing old stars,
while the R clusters are primarily forming in regions of shocked gas at the 
interface between the two colliding galaxies where there are very few 
existing old stars.

\subsection{\ha Line Emission (6563~\AA)}

The \ha recombination line is produced in
HII regions, either photoionized by hot stars or collisionally ionized by
interstellar shocks. \ha emission is often treated as an effective 
indicator of massive star formation. 
The \ha image of the Antennae we analyze here was obtained with WFPC2 
using the F658N narrow-band filter (Whitmore et al. 1999) and was
continuum subtracted to obtain the flux of pure \ha line emission. 
Both diffuse structures and highly concentrated blobs are traced
by the \ha line emission.
The bright concentrations of clusters are often surrounded by filaments, 
which appear to be products of stellar winds or supernovae remnants.

Figure 11 shows the star clusters overlaid on a gray-scale \ha
image of the Antennae. We do not plot a contour map of \ha emission 
because it is difficult to show the many fine structures within the Antennae.
The R clusters follow the overall \ha flux distribution, but they are rarely 
coincident with the strongest \ha peaks, either in the western loop of 
NGC~4038 or in the overlap region, where a large fraction of the \ha flux is 
detected. Instead, they are generally offset from these peaks. 
This may be because some ionizing UV photons from the R clusters are 
absorbed by dust rather than by gas.
The B1 clusters are associated with nearly all the \ha concentrations, 
demonstrating that \ha emission is an excellent indicator of recent 
cluster formation. The B2 clusters, on the other hand, are not as closely 
associated with the most prominent \ha peaks.
This is to be expected since the ionizing radiation from a stellar population
decays dramatically after $\sim 10$~Myr (Leitherer et al. 1999). 

An exception to the overall positive correlation between \ha concentrations 
and the B1 clusters is in the eastern star formation region. Here the \ha 
emission is low and there is relatively little absorbing gas or dust.
The presence of large number of B2 clusters in this region suggests the 
possibility that a burst of star formation some $10-100$~Myr ago 
blew away some of the surrounding gas.
Some support for this latter speculation is provided by the arc-like shape 
in the CO~(1-0) contours which envelopes the northwestern side of this region 
(see Figure~7) and the indentation in the west side of the region in the 
far-infrared map (see Figure~8).

The cross-correlation functions between star clusters and \ha line emission
is shown in Figure~11$d$.
The correlation for the B1 clusters is much stronger
than the correlation for the R clusters or the B2 clusters. 
Note the very high value of $\xi(r)$ for the B1 clusters.
The half-maximum width of $\xi(r)$ for the B1 clusters is $\approx 2.2\arcsec$, 
which corresponds to $\approx 200$~pc. 
This is roughly the distance traveled in 8~Myr at the typical speed 
of 25~\kms\ for stellar winds in the Antennae (Whitmore et al. 1999), and
is consistent with the typical size of giant HII regions surrounding the 
clusters (e.g., knots~G, B, K, M, and S; see enlargements in Figures~7 and 8 
of Whitmore et al. 1999).
Moreover, the dramatic difference in $\xi(r)$ between the B1 and B2 samples
gives us confidence in the selection of different age populations using the 
reddening-free $Q$ parameters.
Finally, $\xi(r)$ for the R clusters is low, and has a peak 
at around $5\arcsec$ ($500$~pc), reflecting their off-peak locations away
from the bright \ha concentrations. 

\subsection{Far-Ultraviolet Emission ($\sim 1500$~\AA)}

Far-Ultraviolet (FUV) emission effectively traces young massive 
O stars while they are hot, provided stellar winds 
(and possibly supernovae) have blown away the surrounding dust.
FUV observations can thus potentially reveal young blue stellar populations
at an age of a few million years.

An FUV image of the Antennae was obtained with the Ultraviolet Imaging 
Telescope ({\it UIT}) during the Astro-2 mission and reported by 
Neff et al. (1997). The spatial resolution and pixel spacing of the 
digital image are $3\arcsec$ (FWHM) and $1.14\arcsec$, respectively. 
The total flux of the Antennae in the FUV band is
$\sim 2.9\times 10^{-13}$ergs s$^{-1}$ cm$^{-2}$\AA$^{-1}$. 
Most of the FUV flux is from NGC~4038 and the eastern edge of the overlap 
region. Only a small fraction is from the nucleus of NGC~4039.
The data we analyze were taken from the {\it UIT} archive at the STScI.

Figure 12 shows the distribution of star clusters on the {\it UIT} 
image. 
The R clusters show little coincidence with the FUV image, while 
the B1 clusters match well with almost all the structures, 
as labeled in Figure~12$a$. 
The B2 clusters coincide with some of the major features, such as the eastern
star formation region (ref. \S5.4) and the western loop of NGC~4038. 
The co-existence of the B1 and B2 clusters in some of these features suggests
a reasonably continuous formation of clusters during the past $\lesssim 
100$~Myr. Moreover, UV flux may linger for as long as $100$~Myr after 
the onset of star formation according to stellar population synthesis models 
(e.g. Leitherer et al. 1999).
While the B1 clusters are clearly sources of FUV emission, 
the R clusters are too deeply embedded in dust for much of the FUV
radiation to escape.
These asosciations are reflected in the cross-correlation
functions in Figure~12$d$, with weak $\xi(r)$ for the R clusters but
strong $\xi(r)$ for the B1 and B2 clusters. The cross-correlation function 
for the R clusters has a peak at $\approx 7\arcsec$, 
similar to the results for the I-band and \ha.

\subsection{Soft X-ray Emission ($0.1-2.5$~keV)}

Soft X-ray emission can originate from several sources: binary stars, 
supernova remnants (hot gas behind shocks), and the diffuse hot interstellar 
medium around young stars. 
It can also result from gas heated by collisional shocks with velocities
around $100-200$~\kms. 

X-ray observations of the Antennae were obtained with the {\it ROSAT} 
High Resolution Imager (HRI) in 1994 and 1995 and reported by 
Fabbiano et al. (1997). 
The HRI energy range is $0.1-2.5$~keV and the spatial resolution is 
$\approx 5\arcsec$. 
The Antennae show intricate structures in X-rays, including both extended 
regions and possible filaments where the emission peaks are associated
with HII regions. 
Fabbiano et al. (1997) suggested that most of the observed emission is 
from a variety of X-ray sources, such as binaries and supernova remnants, and 
from a diffuse hot ISM. 
The image we use in the following was processed from Figure~3$c$ of 
Fabbiano et al. (1997).

Figure 13 shows the distribution of star clusters overlaid
on the contours of the  {\it ROSAT} HRI observations.
The R clusters have a relatively poor association with
the X-ray emission, which is absent in most of the dusty overlap region. 
This could be due to absorption (mainly by heavy elements).
The B1 and B2 clusters, on the other hand,  appear to be 
associated with many of the X-ray features.
Fabbiano et al. (1997) reported the coincidence of some radio sources
with high X-ray emissivity in the western loop of NGC~4038 
and the galactic nuclei. 
Figure~13$d$ shows the cross-correlation functions for the clusters
and X-ray emission. As anticipated, $\xi(r)$ is large and almost identical 
for the B1 and B2 clusters, but is much weaker for the R clusters. 

The B1 and B2 clusters are found to be closely associated with X-ray emission 
in the eastern star formation region where the only other good correlation 
is with the FUV emission.
Young clusters coincide exactly with the point source X-11 
(cf. Fabbiano et al. 1997), where radio and infrared emission is weak. 
This region probably contains clusters with a range of ages.
Unlike X-11, its eastern companion X-12 has no associated clusters.
As noted by Fabbiano et al. (1997), there is evidence of a super shell 
surrounding the clusters in the eastern star formation region. 
As Figure~7 shows, molecular gas surrounds the clusters on the north and 
west sides of this region. Radio and far-infrared emission also avoid 
this region, as shown in Figures~5 and 8.
A possible explanation is that these clusters, which formed $10-100$~Myr
ago, have experienced supernova explosions, and that their remnants 
are the source of the X-ray emission. 
The gas driven to the east by the supernova explosions may have 
confronted less gas than to the west, hence traveling farther
from the clusters.

\subsection{Robustness of $\xi(r)$}

Finally, we discuss how the cross-correlation functions
may be affected by excluding unrepresentative regions from the sample.  
The most prominent example is the eastern star formation 
region, where there exist more than $100$ B1 and B2 clusters but few R 
clusters.  We repeat the calculations of $\xi(r)$ after excluding this 
region and then compare with the earlier results. As shown in 
Figure~14, the difference is generally small or modest ($<25\%$ at small 
$r$) for the B1 sample, while it is more noticeable for the B2 sample 
(e.g., for HI, radio continuum, and $60\micron$). We note that the changes
do not alter the relative rank order of the cross-correlation functions
for the R, B1, and B2 samples, even in this extreme case.

\section{Velocity Fields and Star Clusters}

Velocity fields in the ISM provide additional information about 
the formation and evolution of star clusters. On the one hand, the merging 
of two galaxies involves the global motion of gas, which may have 
caused shocks and triggered star and cluster formation.
On the other hand,  feedback (photoionization, stellar winds and 
supernova explosions, etc.) 
from newly-formed clusters inputs energy into the ISM and influences the
local motions within galaxies.
Therefore, velocity fields may contain information concerning both the cause
and the effects of the formation of young star clusters.

In the following, we explore the relationship between the clusters and the
velocity fields of the HI and HII in the Antennae galaxies. 
In addition to overlaying the clusters on the velocity maps, 
we also calculate the cross-correlation function between
the star clusters and the velocity gradients and the velocity dispersions.
The regions without contours are those with no velocity information available,
and are not considered in the calculation of cross-correlation functions.

\subsection{HI Velocity Field}

The HI (21~cm) mean velocity and velocity dispersion maps we analyze here
were obtained with the VLA with a velocity resolution of $5.21$~\kms\
(Hibbard et al. 2001), 
and are from the same observations as the flux map discussed in \S5.1.

Figure 15 shows the clusters overlaid on the contours of the HI 
intensity-weighted velocity map.
Despite the  somewhat distorted disk structures, the atomic gas shows 
a relatively smooth velocity distribution over the two colliding galaxies.
High velocity gradients exist in only two regions where the velocity contours 
are dense. One is located around ${\rm R.A.} = 12^h01^m55.0^s$ and 
${\rm Decl.}=-18\arcdeg52\arcmin00.0\arcsec$, where the overlap region and 
the base of the NGC~4038 tail are superposed. The other is in the southern 
part of the overlap region. 
Overall, the star clusters do not reside preferentially in regions
of high or low velocity gradient, as is reflected in Figure~$15d$.

Figure~16 shows the clusters overlaid on the contours of the intensity-weighted
velocity dispersion map. As seen in panel~$d$, there are weak but 
non-negligible statistical associations for the R ($2\sigma$) and the B1 
($5\sigma$) samples. 
While none of these correlations are compelling by themselves,
the fact that they increase in the age sequence B2, B1, R in a similar
way as found in Figures~4, 5, 7, and 8, provides additional support for 
their reality.

\subsection{\ha Velocity Field}

The \ha velocity map we analyze was obtained using a Fabry-Perot 
interferometer on the ESO 3.6m telescope, as reported by Amram et al. (1992). 
The spatial resolution is $0.91\arcsec$ and the spectral resolution is 
$\approx 16$~\kms. 
The processed data were kindly provided by M. Marcelin and P. Amram. 

Figure 17 shows the clusters with the contours of the \ha velocity field 
superimposed.
The \ha velocity field in the northern part of NGC~4038 shows a relatively
smooth pattern similar to that of the atomic gas,
while those in the southern part of NGC~4038 and in
the overlap region show many loops and disturbed structures. 
Interestingly, the three major loops in the overlap region are exactly 
centered on the knots~B, C, and D, as labeled in Figure~$17a$, with the central 
line-of-sight velocity smaller than those of the surrounding annuli. 
The contour loops are indicative of super shells blown by the newly formed 
star clusters in these regions. 
We examine knot~D in more detail below.
Note that similar velocity loops are not seen in other regions, such
as knots~J, K, G, S, and T, even though these knots also have \ha super
shells around them (possibly due to the heavy dust obscuration in knots~J
and K, and the greater ages in knots~G, S, and T).

The \ha velocity near knot~D  is $1457$~\kms. 
The velocity gradient is nearly the same on the north, west, and south sides 
of the knot, with the velocity increasing by $38$~\kms\ at a distance of 
$5.2\arcsec$, corresponding to $0.48$~kpc.
For comparison, the central velocities of knots~B and C are $1485$~\kms\
and $1442$~\kms, respectively, and the velocities increase by
$\approx 39$~\kms\ and $\approx 60$~\kms\ at distances of $0.5$~kpc and 
$0.28$~kpc, respectively. 
On the fourth (east) side of knot~D, a blowout has apparently occurred.
As shown in Figure~7, knot~D resides near the edge of the molecular cloud
complexes in the overlap region. 
Thus, the eastward opening of the \ha velocity contours is probably due 
to the relatively dilute gas in this direction.
It is interesting to note that the \ha velocity contours around knot~D
exactly match the indentation in the HI flux map in Figure~4, showing
the lack of atomic gas to the east.
In addition, knots~B, C, and D have velocities consistent with 
their interstellar environment ($\approx 1500$~\kms). 

Clearly, there is strong evidence for feedback from the B1 clusters in the 
overlap region. The outflows from these knots have velocities slightly 
larger than the typical velocity of $25-30$~\kms\ in knots~S and K 
(Whitmore et al. 1999), and may inject a large amount of energy into the 
local ISM.  Moreover, the velocity of the straight dense contours just 
north of knots~C and D increases from 1495~\kms\ to 1564~\kms\ within a 
distance of $\approx 2.4\arcsec$ (0.22~kpc) corresponding to a gradient of 
about 314~\kms~kpc$^{-1}$. 
We speculate that the outflows in this region may be responsible 
for triggering the formation of some of the R clusters. 

Statistically, there is no obvious correlation between the clusters and the 
\ha velocity field for the R and B2 samples in Figure~$17d$. 
There is a negative correlation ($10\sigma$) for the B1 sample, 
due to the flatter velocity gradients in the centers of the super shells  
than in their surrounding velocity loops. 
This negative correlation survives the test of regions when measuring
the correration function as long as knots B, C, and D are included.

\section{The Schmidt Law in the Antennae}

The Schmidt law, which relates the star formation rate and the surface 
density of gas in the form $\Sigma_{\rm SFR}=A \Sigma_{\rm gas}^N$,
has been empirically tested for many galaxies. 
The index $N$ is found to be $1.4\pm 0.15$ over a large sample 
including a variety of galaxies from ``quiescent'' disk 
galaxies to starbursts (Kennicutt 1998).
It is also noted that, for star formation to commence, the density of gas 
must reach a threshold of $\sim 10^{20}$~H cm$^{-2}$ 
($\sim 1~\Mo~{\rm pc}^{-2}$), which appears to be consistent with the 
critical density of gravitational instability in spiral disks (Kennicutt 1989). 
Given the prominent star and cluster formation in the Antennae, it would be 
interesting to check if the Schmidt law is also valid
within this pair of merging galaxies.

We explore the relationship between the star formation rate and gas 
density at various locations within the Antennae galaxies. 
First we grid the central disk into small regions and calculate 
the total flux of \ha as well as the total gas content in each cell
by summing up both the atomic and molecular gas. 
To obtain the mass of the atomic hydrogen from the 21~cm flux we adopt 
the conversion formula 
$M_{\rm HI}/\Mo=2.36\times 10^5 D^2 \int S_{\nu} dv$,
where $D$ is the distance in Mpc and $S_{\nu} dv$ is in Jy~\kms
(Rohlfs \& Wilson 2000, their equation 12.57). 
We obtain a total mass of $3.95\times 10^9~\Mo$ in atomic hydrogen
including the tidal tails, consistent with the value reported by 
Hibbard et al. (2001). 
To obtain the molecular gas content from the CO~(1-0) line emission we use
the same conversion factor $N({\rm H_2})/I_{\rm CO}
=3.0 \times 10^{20} {\rm cm^{-2}(K~km~s^{-1})^{-1}}$ as in \S5.3,
and obtain a relation $M({\rm molecular gas})/\Mo 
= 1.60\times 10^4 (D/{\rm Mpc})^2 S_{\rm CO}/{\rm Jy~km~s^{-1}}$.
This includes an additional correction of $36~\%$ for heavy elements 
(cf. Rohlfs \& Wilson 2000).
We thus obtain a total molecular mass of $1.44\times 10^{10}~\Mo$ within the
central disk of the Antennae, similar to the value reported by 
Gao et al. (2001). 
Finally, we convert the flux of \ha line emission (continuum-subtracted)
into the star formation rate using the relation  
$\Sigma_{\rm SFR}(\Mo{\rm yr}^{-1})= 
7.9\times 10^{-42}L({\rm H\alpha}) ({\rm ergs~s}^{-1})$ (Kennicutt 1998;
see references therein for the calibration). 

Wilson (1995) found that the CO-to-H$_2$ conversion factor varies with
the metallicity in the host galaxy.  Since NGC~4038 and NGC~4039 have
luminosities and morphologies similar to giant spiral galaxies, it is
likely that their metallicity is similar to that of the Milky Way and M31.
Hence, we will adopt the empirical conversion factor listed above
throughout this section.  We also  note that any variation in the metallicity
will  shift the data without altering the slope of the 
$\Sigma_{\rm SFR}-\Sigma_{\rm gas}$ relation. Other factors that
may cause variations in the conversion factor include the size of
molecular clouds, the temperature of the ISM, and the optical depth of
the CO line. Moreover, there may be complications from the dissociation and
consumption of molecular gas due to the starburst in the Antennae.

We correct for extinction in the \ha flux by assuming a uniform value 
over the whole galaxy. It is not possible to correct for each individual
cluster using the $QQ$ analysis (see \S2), since we only have 
useful $U$ measurements (hence $Q_1$ values) for $37\%$ of the R clusters,
where extinction is most important. However, we can correct for each
of the three samples using the mean values for the clusters for which
we do have $Q$ values. In this way, we obtain $\bar{A}_V\approx 4.7$ for 
the R sample, $1.6$ for the B1 sample, and $0.6$ for the B2 sample, 
resulting in an average of $\bar{A}_V\approx 1.6$ for all clusters.
The mean extinction estimated in this way should be regarded as the lower limit
since it may be biased toward the least extincted clusters (i.e., 
we tend to miss clusters with the highest values of $A_V$). 
Another method is to estimate the extinction from the measured
H$\alpha$/H$\beta$ line ratio that was reported by Keel et al. (1985).
We obtain a value of $A_V\approx 2.8$ with the intrinsic H$\alpha$/H$\beta$ 
ratio of $2.86$ (assuming a Galactic extinction law and $T_e=10^4$~K, 
Caplan \& Deharveng 1986). 
Since the line ratio is affected by absorption lines from young stars, 
this estimate should be regarded as an upper limit.
Thus, we find $A_{\rm H\alpha}$ to be within the range $1.3-2.3$,
after interpolating between $A_V$ and $A_I$.
In the following we will use the middle value of $1.8$. 
Better estimates of extinction can eventually be made as more infrared
spectra become available.

The relation between $\Sigma_{\rm SFR}$ and $\Sigma_{\rm gas}$
in the Antennae galaxies is shown in Figure~18. 
Figure~$18a$ shows the result from a grid size of $11.4\arcsec\times 11.4\arcsec$
($1.06\times 1.06$~kpc), comparable to the resolution of the HI observations. 
The data show significant scatter; the Spearman coefficient is $0.72$ 
for the total gas density.
Panel~$b$ shows the result from a larger grid size of 
$28.5\arcsec\times 28.5\arcsec$ ($2.65\times 2.65$~kpc). 
The Spearman coefficient is $0.89$ for the total gas density.
We fit the data by using first $\Sigma_{\rm SFR}$ then $\Sigma_{\rm gas}$ 
as the independent variable, and then using the averaged relation as our
final result, as shown by the dotted lines in Figure~18.
The slopes of the data in panel~$a$ and $b$ are $N=1.20\pm 0.23$, and $1.38\pm
0.20$, respectively, and are consistent with the Schmidt law with 
$N=1.4\pm 0.15$ within the $1\sigma$ error, especially in the case of the 
larger grid size. 
Repeating the analysis using the Wilson et al. (2000) CO intensity
map rather than the BIMA map results in correlations with similar slopes.
A possible explanation for the poorer relation on smaller grid scales
is that below a certain size limit the young clusters may alter their 
interstellar environment by depleting the local molecular and atomic 
hydrogen gas. Hence, the Schmidt law may be modified at scales smaller
than the range of the feedback effects. 

The agreement with the Schmidt law is somewhat surprising, given the large
values of $A_V$ in the overlap region and the fact that we are probably
missing some obscured clusters (see \S5.2). We have tested 
the uniformity of this result and the dependence on our simplistic 
treatment of $A_V$ by repeating the calculations using only 
the northern or western half of the maps. Based on the $QQ$ analysis, 
the extinction in these regions is much smaller than in the overlap region. 
The tests show that the slopes for a $28.5\arcsec\times 28.5\arcsec$ grid 
vary from $1.2$ to $1.5$, with the slope of $1.4$ being always within 
the $1\sigma$ uncertainty. The slopes for a smaller grid of 
$11.4\arcsec\times11.4\arcsec$ are between $1.0$ and $1.1$,
consistent with the value $1.4$ within the $2\sigma$ uncertainty.

We also present the relationship between the cluster formation rate 
($\Sigma_{\rm CFR}$) and the gas content in the Antennae in Figure~19. 
The $\Sigma_{\rm CFR}$ is defined as the number density of the R and B1
clusters divided by $8$~Myr, roughly their average age.
The slopes of the fits are $N=1.37\pm 0.51$ and $N=1.53\pm0.57$
for the upper and lower panels, respectively.
The relation between $\Sigma_{\rm CFR}$ and gas content has
more scatter than that between $\Sigma_{\rm SFR}$ and gas content.
This might be expected due to small number statistics (e.g., the bottom
row of points has $n=1$), the possibility of missing some embedded clusters due 
to obscuration, and incompleteness at the faint end.
However, the similarity in the slopes of these relations suggests that 
the Schmidt law is also a valid description of cluster formation in
merging galaxies.

We can also estimate how long it would take before all the gas in the Antennae 
was consumed by star formation at the current SFR (ignoring possible inflow
and outflow). This is sometimes called the 
``Roberts time'', $\tau_R$, and is defined as 
$M_{\rm gas}/{\rm SFR}$ (cf. Kennicutt, Tamblyn, \& Congdon 1994).
The overall star formation rate is $4.0~\Mo$~yr$^{-1}$, derived directly 
from the total \ha flux of
$1.13\times 10^{-11}$~ergs~s$^{-1}$~cm$^{-2}$ within the disk of 
the Antennae before correction for extinction. 
The star formation rate is $21.0~\Mo$~yr$^{-1}$ if $A_{{\rm H}\alpha}=1.8$ 
is adopted. This is larger than the value of $5~\Mo$~yr$^{-1}$
estimated by Stanford et al. (1990), mainly due to the higher \ha line flux 
detected with the {\it HST}.
For comparison, the current star formation rate of the Milky Way is about
3~$\Mo$~yr$^{-1}$ (Mezger 1978; Scoville \& Good 1989), 
and that for M82 is about $2~\Mo$~yr$^{-1}$ (e.g., Condon 1992).
Taking a total gas mass of $\approx 1.5\times 10^{10}~\Mo$ in the Antennae,
we obtain $\tau_R \approx 700$~Myr, which 
is comparable with the merger age of $\sim 500$~Myr
(cf. Mihos, Bothun, \& Richstone 1993), but is shorter than the typical Roberts
time for spiral galaxies ($\sim 3$~Gyr, Kennicutt et al. 1994).
These time scales indicate that star formation in the Antennae has been
enhanced by the merger event. 

\section{Summary}

The recent availability of state-of-the-art maps of the Antennae galaxies
(NGC~4038/9), covering almost the whole wavelength range from radio to X-ray,
has made it possible for the first time to study the relationship between 
young star clusters and their interstellar environment.
The Antennae system provides the ideal laboratory for this study since it is both
the youngest and nearest of Toomre's (1977) prototypical mergers. 
We use these new data to perform a comprehensive study of the 
correlation between young star clusters of different ages
and their interstellar environment in the Antennae.

Using the {\it UBVI} images from {\it HST} WFPC2 observations, 
we identify three age groups of young star clusters in the Antennae: 
   red clusters (the R sample, $t\sim 5$~Myr), 
   young bright clusters (the B1 sample, $3\lesssim t \lesssim 16$~Myr), 
   and older bright clusters (the B2 sample, $16\lesssim t \lesssim 160$~Myr).
The observations in the other wavelength bands used in this study include
21~cm line emission (HI), radio continuum (6-cm), CO (1-0), far-infrared 
(60\micron), mid-infrared (15\micron), \ha line, FUV ($\sim 1500$~\AA), and 
soft X-ray emissions. 
Velocity maps are available in 21~cm, CO (1-0), and \ha lines. 
We first calculate the autocorrelation functions for the star clusters, 
and then study their cross-correlation with the flux maps 
and the velocity fields. 
We note that some observations at longer wavelength are restricted 
to lower resolution (e.g., $1.58$~kpc for the far-infrared emission and 
$1.06\times 0.69$~kpc for the HI 21~cm line 
emission), and thus some important small-scale information is not available.
In addition, selection effects caused by isolating the R sample based on 
color potentially affect the interpretation of some of the correlations 
(e.g., the $60\micron$ and $15\micron$ observations, which correlate with the 
presence of dust). This is less of a problem for the B1 and B2 clusters which
are selected based on reddening-free $Q$ parameters rather than their
apparent colors.

Our main results are as follows:

1. We find that star clusters form in a clustered distribution. 
Their two-point autocorrelation function is a power law with index
in the range $-1.0$ to $-0.8$. 
In contrast, the correlation function for the sample of candidate young  
massive stars is much weaker, with a power-law index $-0.4$.
Moreover, the scale of $\sim 1$~kpc, beyond which the power-law correlation 
functions for the clusters drop dramatically, is consistent with the typical
size of giant molecular cloud complexes identified within the Antennae galaxies.
These molecular cloud complexes typically have masses of a few times $10^8~\Mo$
(Wilson et al. 2000).

2. We find that the cross-correlations between the star clusters and 
the flux maps vary with wavelength.
The R clusters are closely related to long-wavelength radiation,
i.e. emission in the 21~cm line, radio continuum, the CO (1-0) line,
the far-infrared and mid-infrared. 
They are poorly correlated with emission that has wavelength shorter than 
near-infrared, apparently a consequence of obscuration by dust.
The B1 and B2 clusters, however, are more closely related to short-wavelength
radiation.
The B1 clusters are the main sources of the FUV and \ha emission,
and are associated with many of the X-ray and I-band features.
The B2 clusters, while largely following the behavior of the B1 clusters in 
the correlation with flux in these bands, have ceased to emit many \ha
photons, but are associated with X-ray and I-band emission. 
The variation of correlations in the different wavelengths is 
consistent with the gradual dissociation of clusters from their formation 
environment: first by heating the dust, then by ionizing the gas, and 
finally by expelling the surrounding materials through stellar winds 
and supernova explosions.

3. The tight correlations between the R clusters and the 
HI 21~cm line  and CO~(1-0) line emission indicate that gas content, 
especially molecular gas, is closely associated with cluster formation.
Many of the R clusters are spatially coincident with giant molecular cloud 
complexes in the overlap region and in the western loop of NGC~4038.
On the other hand, gas concentrations alone, even molecular, do
not guarantee that clusters will form.
For example, few R clusters are found in the CO (1-0) flux concentrations 
and atomic gas just south of the nucleus of NGC~4039, in the southeast
extension (see Figure~7), and in the tails, where HI is abundant.

4. There is some evidence for weak correlations between the locations of 
young star clusters and the velocity dispersion of atomic gas
($5\sigma$). 
The most prominent ongoing cluster formation in the southern overlap region
is coincident with chaotic velocity fields
with line-of-sight velocity gradients of $\sim 100$~\kms~kpc$^{-1}$.
This region of high velocity gradient is also the location with the largest 
velocity dispersion in molecular gas.  Wilson et al. (2000) suggested that 
it is the site where several giant molecular clouds are colliding. 
Shocks may have triggered the formation of the clusters in this region.
However, collisions among molecular clouds do not seem to account for
most of the R clusters, which are formed in other, less chaotic locations.

5. There is strong evidence for feedback by young clusters. 
We find prominent \ha super shells surrounding bright cluster complexes
of the B1 sample, especially in the overlap region.
These outflows have typical velocities of $25-30$~\kms\ or larger
and inject a large amount of energy into the local ISM.
They may be responsible for triggering the 
formation of some of the R clusters (e.g., north of knots~C and D 
where the velocity gradient in ionized gas is large).
We thus infer that the Antennae are a complicated system where different
triggering mechanisms may be at work.

6. We find that, in the Antennae, the relationship between the star 
formation rate and the surface density of gas can be described by the 
Schmidt law with index $N=1.4\pm 0.15$. 
The cluster formation rate obeys the same relationship but with more scatter.
However, the Schmidt law may be modified at scales smaller than
$\approx 1$~kpc, where feedback effects become important.

7. We estimate the total observed flux of \ha line emission to be
$1.13\times 10^{-11}$~ergs~s$^{-1}$~cm$^{-2}$ in the central disk of
the Antennae galaxies.  The star formation rate is $\approx 20~\Mo$~yr$^{-1}$, 
if $A_{{\rm H}\alpha}=1.8$ is adopted. 
This yields a gas consumption time (i.e., the Roberts time) of
$\tau_R\approx 700$~Myr, which is comparable with
the merger time but is smaller than that of a typical spiral galaxy,
suggesting a major enhancement of star and cluster formation due to the 
merger event in the Antennae. 

\acknowledgements 
We thank John Hibbard for providing the VLA 21~cm intensity and velocity maps 
and Susan Neff and Jim Ulvestad for the radio (6~cm and 3.5~cm) data 
prior to publication.
We also thank Rhodri Evans for the KAO $60\micron$ data,
Felix Mirabel for the ISOCAM map, 
Michel Marcelin and P. Amram for the \ha velocity map, and
Christine Wilson for the CO~(1-0) intensity map.
We thank Ron Allen and an anonymous referee for helpful comments 
that have improved the paper.

\newpage
\figcaption{
%figure 1 
Comparison of the R clusters and stellar population synthesis models 
in the color-color diagram and the reddening-free $Q$-parameter diagram.
Only the R clusters with errors in $U-B$ and $B-V$ less than 1 magnitude
are shown. The solid and dotted lines represent the BC96 and Starburst99 
models with solar metallicity, respectively. 
The numbers label the logarithmic age in years for the models. 
The arrow in the upper panel 
shows a standard reddening vector with 1 magnitude of extinction. 
The cross in the lower panel shows the error-weighted average
position of the objects with $Q_1<-0.7$. 
Note that most of the R clusters have an age of $\sim 5$~Myr.
}

\figcaption{
%figure 2 
Spatial distribution of the R ($t\sim 5$~Myr), 
B1 ($3\lesssim t \lesssim 16$~Myr), 
and B2 ($16 \lesssim t \lesssim 160$~Myr) clusters and  
candidate young stars ($-9<M_V<-6$). 
The symbols in the right panels are made smaller for clarity.
The ellipses label some regions of interest while the letters in panel~$b$ 
mark the bright knots of star clusters. 
The upper and lower crosses (``+'') label the nuclei of 
NGC~4038 and NGC~4039, at ${\rm R.A.} = 12^h01^m53.0^s$,
${\rm Decl.}= -18\arcdeg52\arcmin01.9\arcsec$, and
${\rm R.A.} = 12^h01^m53.5^s$, 
${\rm Decl.}= -18\arcdeg53\arcmin09.8\arcsec$ (J2000.0),
respectively, measured from radio continuum (6~cm and 3.5~cm) maps.
The straight lines show the WFPC2 window.
}

\figcaption{
%figure 3 
Two-point autocorrelation functions $\xi(r)$ of clusters and candidate 
young stars. 
The dashed curve is for the R sample, the solid curve is for the B1 sample, 
and the dot-dashed curve is for the B2 sample. 
The dotted curve represents the candidate young stars.
The vertical lines show the uncertainties in $\xi(r)$.
The top panel shows the autocorrelation function $\xi(r)$ without any 
consideration of the large-scale structure of the galaxy. 
The middle panel shows $\xi(r)$ between the objects and their smoothed 
density distribution (using a Gaussian mask with ${\rm FWHM}=3$~kpc). 
The bottom panel shows $\xi(r)$ with the large-scale structure of the 
galaxy subtracted.  Note the flatness in $\xi(r)_s$ for small $r$.
}

\figcaption{
%figure 4: clusters vs. HI 
Association between 21~cm HI flux (contours) and 
star clusters (crosses). 
Panels~$a$, $b$, and $c$ show the locations of the R clusters 
($t\sim 5$~Myr), B1 clusters ($3\lesssim t \lesssim 16$~Myr) and 
B2 clusters ($16 \lesssim t \lesssim 160$~Myr), respectively.
The HI contour levels are 4.3, 6.5, 9.7, 14.5, 21.8, 32.7, 49.1, 73.6,
110.5, 165.7$\times 10^{-3}$~mJy~\kms.
The ellipse at the lower-right corner of panel~$a$ shows the beam size of 
the contour map.
Panel~$d$ shows the cross-correlation functions between the star clusters and 
HI flux, with the large-scale structure of the galaxy subtracted.
The three sets of vertical lines at $r\approx1\arcsec, 10\arcsec, 20\arcsec$ 
show the uncertainties in $\xi(r)$.
The small vertical tick near the zero point shows the equivalent
half width at half maximum (HWHM) of the HI beam.
The horizontal axis on the top labels the radius in kpc, assuming a distance
of $19.2$~Mpc for the Antennae galaxies.
}

\clearpage
\figcaption{
%figure 5 
Association between radio continuum (6~cm) flux and star clusters.
The contour levels are  1.7, 2.6, 3.9, 5.8, 8.8, 13.1, 19.7, 29.5, 44.3, 
66.5~$\mu$Jy. 
The notations are the same as in Figure~4.
}

\figcaption{
%figure 6 
Distributions of R clusters, bright B1 clusters, and 
compact radio sources with flux larger than twice the detection limit
identified with radio 6~cm continuum observation  by Neff \& Ulvestad 
(2000, their table~3). 
The size of the open circles approximately represents the resolution of 
the radio continuum observation.
The boxes label the knots~B, C, D, E, F, J, and K.
The reference point at (0, 0) is
${\rm R.A.} = 12^h01^m52.97^s$ and
${\rm Decl.}= -18\arcdeg52\arcmin08.29\arcsec$ (J2000.0). 
}

\figcaption{
%figure 7 
Association between CO~(1-0) flux and star clusters.
The contour levels are 2.9, 4.3, 6.5, 9.8, 14.6, 22.0, 33.0, 49.4,
74.1, 111.2$\times 10^{-3}$~Jy~\kms.  
The label WS\#80 in panel~$a$ corresponds to the red object \#80 listed in
Whitmore \& Schweizer (1995).
The other notations are the same as in Figure 4.
}

\figcaption{
%figure 8 
Association between far-infrared (60\micron) flux and star clusters.
The contour levels are 0.8, 1.3, 1.8, 2.3, 2.8, 3.3, 3.8, 4.3, 4.8, 5.3, 5.8, 
6.3, 6.8, 7.3, 7.8 ADU (Note: unlike other figures, this is a linear scale). 
The three sets of vertical lines at $r\approx1\arcsec, 10\arcsec, 20\arcsec$ 
show the uncertainties in $\xi(r)$.
The notations are the same as in Figure~4.
}

\figcaption{
%figure 9 
Association between near-infrared (15\micron) flux and star clusters.
The contour levels are 0.024, 0.037, 0.055, 0.082,  0.12,  0.18,
0.28, 0.42, 0.62,  0.94~mJy.  
The notations are the same as in Figure 4.
}

\figcaption{
%figure 10 
Association between I-band flux and star clusters.
The contour levels are 0.50, 0.75, 1.13, 1.69, 2.54, 3.81,
5.72, 8.58, 12.86, 19.30 $\times 10^{-18}$~ergs s$^{-1}$ cm$^{-2}$\AA$^{-1}$. 
The notations are the same as in Figure~4.
}

\figcaption{
%figures 11 
Association between the \ha line flux and star clusters.
The \ha map is shown in gray scale.
The notations are the same as in Figure~4.
}

\figcaption{
%figure 12 
Association between FUV flux and star clusters.
The contour levels are $0.8, 1.2, 1.8, 2.7, 4.0, 6.0, 9.0, 13.5, 20.3, 
30.4~\times 10^{-18}$~ergs~s$^{-1}$~cm$^{-2}$~\AA$^{-1}$.
The letter labels in panel~$a$ correspond to the bright cluster regions 
categorized by Whitmore et al. (1999).
The other notations are the same as in Figure 4.
}

\figcaption{
%figure 13 
Association between X-ray flux and star clusters.
The contour levels are 15,  29,  43,  57,  71 (in arbitrary units). 
The letter labels in panel~$a$ correspond to regions categorized by
Fabbiano et al. (1997).
The other notations are the same as in Figure~4.
}

\figcaption{
%figure 14 
Comparison of $\xi(r)$ with (thick lines) and without (thin lines)
clusters from the eastern star formation region. 
The solid lines represent the B1 sample, while the dot-dashed lines
represent the B2 sample.
The dashed line represents $\xi(r)$ for the R clusters (unchanged
since there are few R clusters in this region).
Note that the rank ordering of $\xi(r)$ for the R, B1, and B2
clusters is not altered by the removal of the eastern star formation region.
}

\figcaption{
%figure 15 
Association between the HI line-of-sight velocity and star 
clusters.
The HI velocity contour levels are equally spaced starting at 1450~\kms, 
with an interval of 15~\kms. 
A gray-scale image is overlaid in panel~$a$ to show the relative line-of-sight
velocities, with darker shades representing higher velocities.
Panel~$d$ shows the cross-correlation functions between the clusters and the
velocity gradients. The red object WS\#80 is marked in panel~$a$, while 
knots~B, E, and F are labeled in panel~$b$.
The three sets of vertical lines at $r\approx1\arcsec, 10\arcsec, 20\arcsec$ 
show the uncertainties in $\xi(r)$.
The other notations are the same as in Figure~4.
}

\figcaption{
%figures 16 
Association between HI velocity dispersion and star clusters.
The HI velocity dispersion contour levels are equally spaced, starting at  
5~\kms, with an interval of 5~\kms.
The numbers show the average velocity dispersions in \kms\ within their
corresponding regions.
A gray-scale image is overlaid in panel~$a$ to show the relative line-of-sight
velocities, with darker shades representing higher velocities.
Panel~$d$ shows the cross-correlation functions between the clusters and the 
velocity gradients.
The three sets of vertical lines at $r\approx1\arcsec, 10\arcsec, 20\arcsec$ 
show the uncertainties in $\xi(r)$.
The other notations are the same as in Figure~4.
}

\figcaption{
%figure 17 
Association between the \ha line-of-sight velocity and star clusters.
The notations are the same as in Figure~15.
}

\figcaption{
%figure 18 
Star formation rate ($\Sigma_{\rm SFR}$) versus gas density 
($\Sigma_{\rm gas}$) in different regions of the Antennae galaxies.
The open circles are for atomic gas only, while the 
filled circles are for the sum of atomic and molecular gas. 
$\Sigma_{\rm SFR}$ is converted from \ha flux using a coefficient of $7.9\times 
10^{42}$ (Kennicutt 1998), and
an extinction of $1.8$ magnitudes in the \ha flux (see text).
The dashed lines are the fits to the total gas density (filled circles).
The solid lines represent the Schmidt law $\Sigma_{\rm SFR}= 
(2.5\pm 0.7)\times10^{-4} \Sigma_{\rm gas}^{1.4\pm0.15}$ (Kennicutt 1998). 
Panel~$a$ shows the relation with a grid resolution of 
$11.4\arcsec\times 11.4\arcsec$, 
panel~$b$ with $28.5\arcsec\times 28.5\arcsec$.
}

\figcaption{
%figure 19 
Cluster formation rate ($\Sigma_{\rm CFR}$) versus gas density
($\Sigma_{\rm gas}$) in different regions of the Antennae galaxies.
The open circles are for atomic gas only, while the 
filled circles are for the sum of atomic and molecular gas. 
$\Sigma_{\rm CFR}$ is defined as the density of the R and B1 clusters
divided by the average age of $8$~Myr.
The dashed lines are the fits to the total gas density (filled circles).
The solid lines represent the relation $\Sigma_{\rm CFR}\propto  
\Sigma_{\rm gas}^{1.4}$.
Panel~$a$ shows the relation with a grid resolution of 
$11.4\arcsec\times 11.4\arcsec$, 
panel~$b$ with $28.5\arcsec\times 28.5\arcsec$.
}

% Table 1
\begin{deluxetable}{ccccccccc}
\small
\tablewidth{0pt}
\tablecaption{Red Clusters}
\tablehead{
\colhead{Number} & \colhead{$\Delta$R.A.\tablenotemark{a}} & 
\colhead{$\Delta$Decl.\tablenotemark{a}} &
\colhead{Chip} & \colhead{$M_V$} &
\colhead{$B-V$} & \colhead{$V-I$} & 
\colhead{$\Delta V_{1-6}$\tablenotemark{b}} & 
\colhead{WS95 Number\tablenotemark{c}}}
\startdata
1  &  -7.71  &  -7.14  &  2  &  -8.32  &  1.47  &  2.25  &  1.92 & 356\\ 
2  &  32.28  &  -38.13  &  3  &  -8.14  &  1.96  &  3.00  &  1.86 & 124\\ 
3  &  28.87  &  -56.58  &  3  &  -7.89  &  1.99  &  2.92  &  2.08 & 80\\ 
4  &  31.17  &  -7.63  &  3  &  -7.63  &  --  &  2.85  &  2.08 & 355\\ 
5  &  28.56  &  -45.48  &  3  &  -7.55  &  2.51  &  2.81  &  2.22 & \\ 
6  &  32.82  &  -36.44  &  3  &  -7.51  &  1.94  &  2.58  &  1.64 & 132\\ 
7  &  28.74  &  -45.20  &  3  &  -7.48  &  1.71  &  2.22  &  3.25 & \\ 
8  &  30.33  &  -30.30  &  3  &  -7.24  &  1.88  &  2.16  &  2.75 & \\ 
9  &  23.20  &  -38.08  &  3  &  -7.18  &  2.22  &  3.25  &  2.21 & 125\\ 
10  &  5.46  &  39.93  &  2  &  -7.16  &  2.71  &  2.04  &  1.67 & \\ 
11  &  31.00  &  -7.63  &  3  &  -7.05  &  2.53  &  3.10  &  3.10 & 355\\ 
12  &  26.48  &  -26.38  &  3  &  -6.97  &  1.63  &  2.32  &  2.06 & \\ 
13  &  36.03  &  -33.44  &  3  &  -6.96  &  2.16  &  2.90  &  1.77 & \\ 
14  &  28.44  &  -49.18  &  3  &  -6.93  &  2.07  &  2.19  &  2.56 & \\ 
15  &  21.01  &  -47.91  &  3  &  -6.88  &  0.50  &  2.04  &  1.84 & \\ 
16  &  3.96  &  33.05  &  2  &  -6.85  &  1.84  &  2.36  &  1.67 & \\ 
17  &  -23.42  &  17.24  &  2  &  -6.77  &  1.97  &  2.07  &  2.57 & \\ 
18  &  -0.64  &  2.57  &  2  &  -6.76  &  --  &  2.33  &  1.92 & \\ 
19  &  32.91  &  -30.09  &  3  &  -6.75  &  2.79  &  2.04  &  2.86 & \\ 
20  &  -2.08  &  11.24  &  2  &  -6.72  &  0.97  &  2.24  &  1.77 & \\ 
21  &  2.36  &  33.25  &  2  &  -6.72  &  0.96  &  2.06  &  2.25 & \\ 
22  &  -8.17  &  -16.34  &  3  &  -6.66  &  1.55  &  2.07  &  2.59 & \\ 
23  &  -19.31  &  -17.68  &  1  &  -6.65  &  1.39  &  2.22  &  2.73 & \\ 
24  &  26.27  &  -24.23  &  3  &  -6.65  &  2.70  &  2.19  &  2.69 & \\ 
25  &  24.45  &  -49.07  &  3  &  -6.63  &  0.82  &  2.11  &  2.32 & \\ 
26  &  -9.64  &  33.61  &  2  &  -6.62  &  1.49  &  2.12  &  2.43 & \\ 
27  &  -27.91  &  0.72  &  1  &  -6.61  &  1.64  &  2.12  &  2.78 & \\ 
28  &  17.58  &  -60.77  &  3  &  -6.60  &  --  &  2.08  &  1.69 & \\ 
29  &  12.10  &  -3.24  &  3  &  -6.59  &  2.07  &  2.13  &  2.89 & \\ 
30  &  1.45  &  25.40  &  2  &  -6.56  &  --  &  2.12  &  1.40 & \\ 
31  &  12.17  &  -3.21  &  3  &  -6.56  &  1.27  &  2.06  &  2.66 & \\ 
32  &  -30.38  &  -6.92  &  1  &  -6.54  &  1.58  &  2.14  &  1.61 & \\ 
33  &  -11.19  &  16.84  &  2  &  -6.54  &  --  &  2.31  &  2.00 & \\ 
34  &  26.08  &  -29.41  &  3  &  -6.52  &  2.29  &  2.14  &  2.10 & \\ 
35  &  40.39  &  -7.66  &  3  &  -6.44  &  --  &  2.36  &  2.57 & \\ 
36  &  9.34  &  33.45  &  2  &  -6.43  &  1.77  &  2.05  &  1.32 & \\ 
37  &  -27.99  &  0.85  &  1  &  -6.41  &  0.85  &  2.20  &  2.61 & \\ 
38  &  -2.65  &  28.67  &  2  &  -6.37  &  1.62  &  2.08  &  1.67 & \\ 
39  &  32.00  &  -38.04  &  3  &  -6.37  &  0.67  &  2.79  &  2.95 & 124\\ 
40  &  31.72  &  -1.77  &  3  &  -6.36  &  --  &  2.39  &  2.35 & \\ 
41  &  -19.90  &  -25.76  &  4  &  -6.29  &  --  &  2.01  &  1.81 & \\ 
42  &  -27.63  &  0.64  &  1  &  -6.29  &  1.35  &  2.20  &  2.38 & \\ 
43  &  67.95  &  -19.34  &  3  &  -6.28  &  0.70  &  2.17  &  2.56 & \\ 
44  &  31.73  &  -11.11  &  3  &  -6.27  &  2.18  &  2.74  &  1.83 & \\ 
45  &  66.37  &  -20.12  &  3  &  -6.24  &  2.02  &  2.13  &  2.86 & \\ 
46  &  -39.36  &  -1.37  &  1  &  -6.23  &  --  &  2.20  &  1.88 & \\ 
47  &  33.55  &  -22.04  &  3  &  -6.22  &  --  &  2.19  &  2.53 & \\ 
48  &  -31.59  &  -5.02  &  1  &  -6.22  &  1.04  &  2.06  &  2.30 & \\ 
49  &  37.27  &  -32.52  &  3  &  -6.20  &  1.19  &  2.37  &  0.37 & \\ 
50  &  26.84  &  -16.52  &  3  &  -6.14  &  1.47  &  2.34  &  2.40 & \\ 
51  &  24.15  &  -22.99  &  3  &  -6.11  &  2.06  &  2.09  &  2.01 & \\ 
52  &  32.65  &  -4.88  &  3  &  -6.11  &  2.81  &  2.03  &  0.61 & \\ 
53  &  23.22  &  -2.54  &  3  &  -6.08  &  2.44  &  2.37  &  1.33 & \\ 
54  &  4.22  &  -59.14  &  4  &  -6.08  &  1.18  &  2.42  &  1.13 & \\ 
55  &  39.66  &  -49.98  &  3  &  -6.06  &  2.67  &  2.31  &  2.15 & \\ 
56  &  48.78  &  15.74  &  3  &  -6.02  &  2.26  &  2.17  &  1.63 & \\ 
57  &  -20.04  &  -14.99  &  1  &  -6.00  &  --  &  2.66  &  1.81 & \\ 
58  &  28.26  &  -42.08  &  3  &  -6.00  &  1.65  &  2.26  &  2.17 & \\ 
59  &  6.12  &  22.69  &  2  &  -5.98  &  --  &  2.09  &  1.53 & \\ 
60  &  -25.74  &  -18.11  &  1  &  -5.94  &  --  &  2.11  &  1.79 & \\ 
61  &  42.69  &  -56.09  &  3  &  -5.92  &  --  &  2.47  &  1.17 & \\ 
62  &  29.25  &  -8.45  &  3  &  -5.85  &  1.57  &  2.20  &  1.93 & \\ 
63  &  34.00  &  -44.39  &  3  &  -5.84  &  1.29  &  2.27  &  1.91 & \\ 
64  &  -33.70  &  -15.47  &  1  &  -5.82  &  --  &  2.58  &  1.28 & \\ 
65  &  46.27  &  -44.23  &  3  &  -5.82  &  --  &  2.23  &  1.98 & \\ 
66  &  40.07  &  -21.60  &  3  &  -5.78  &  --  &  2.38  &  1.55 & \\ 
67  &  26.03  &  -24.79  &  3  &  -5.76  &  2.17  &  2.59  &  1.81 & \\ 
68  &  18.73  &  -0.58  &  3  &  -5.73  &  --  &  3.32  &  2.50 & \\ 
69  &  -30.50  &  -15.71  &  1  &  -5.73  &  2.33  &  2.30  &  1.62 & \\ 
70  &  31.70  &  -29.35  &  3  &  -5.67  &  1.38  &  2.71  &  1.96 & \\ 
71  &  -42.57  &  1.88  &  1  &  -5.64  &  0.89  &  2.36  &  2.43 & \\ 
72  &  -41.09  &  -2.09  &  1  &  -5.55  &  --  &  2.60  &  2.23 & \\ 
73  &  36.10  &  -21.07  &  3  &  -5.39  &  --  &  2.95  &  1.52 & \\ 
74  &  25.61  &  -42.25  &  3  &  -5.39  &  1.04  &  2.54 &  1.78 & \\ 
75  &  45.50  &  -33.12  &  3  &  -5.30  &  0.69  &  2.42  &  1.03 & \\ 
76  &  -29.78  &  3.48  &  1  &  -5.24  &  --  &  2.58  &  2.25 & \\ 
77  &  26.15  &  -28.16  &  3  &  -5.18  &  1.29  &  3.15  &  1.10 & \\ 
78  &  -48.24  &  -11.58  &  1  &  -5.12  &  --  &  2.72  &  1.57 & \\ 
79  &  30.40  &  -44.14  &  3  &  -4.94  &  --  &  2.86  &  1.11 & \\ 
80  &  24.94  &  -42.67  &  3  &  -4.88  &  1.04  &  4.12  &  1.76 & \\ 
81  &  15.90  &  -32.21  &  3  &  -4.83  &  1.25  &  2.71  &  0.35 & \\ 
82  &  -20.49  &  -21.53  &  4  &  -4.82  &  --  &  2.81  &  1.91 & \\ 
83  &  22.95  &  -49.74  &  3  &  -4.69  &  --  &  2.97  &  0.19 & \\ 
84  &  -41.36  &  -4.80  &  1  &  -4.37  &  --  &  3.04  &  0.05 & 
\enddata
\tablenotetext{a}{Following the convention used in Whitmore \& Schweizer 
(1995) and 
Whitmore et al. (1999), the coordinates are the offsets in arcseconds from 
${\rm R.A.} = 12^h01^m52.97^s$ and 
${\rm Decl.}= -18\arcdeg52\arcmin08.29\arcsec$ (J2000.0), near the center
of NGC~4038.}
\tablenotetext{b}{Objects with $\Delta V_{1-6} < 1.0$ are likely to be cosmic
rays.}
\tablenotetext{c}{Number assigned by Whitmore \& Schweizer (1995). 
Duplicated numbers are neighboring objects, generally because
of differences in spatial resolution between the cycle 2 and cycle 5 observations.}
\end{deluxetable}

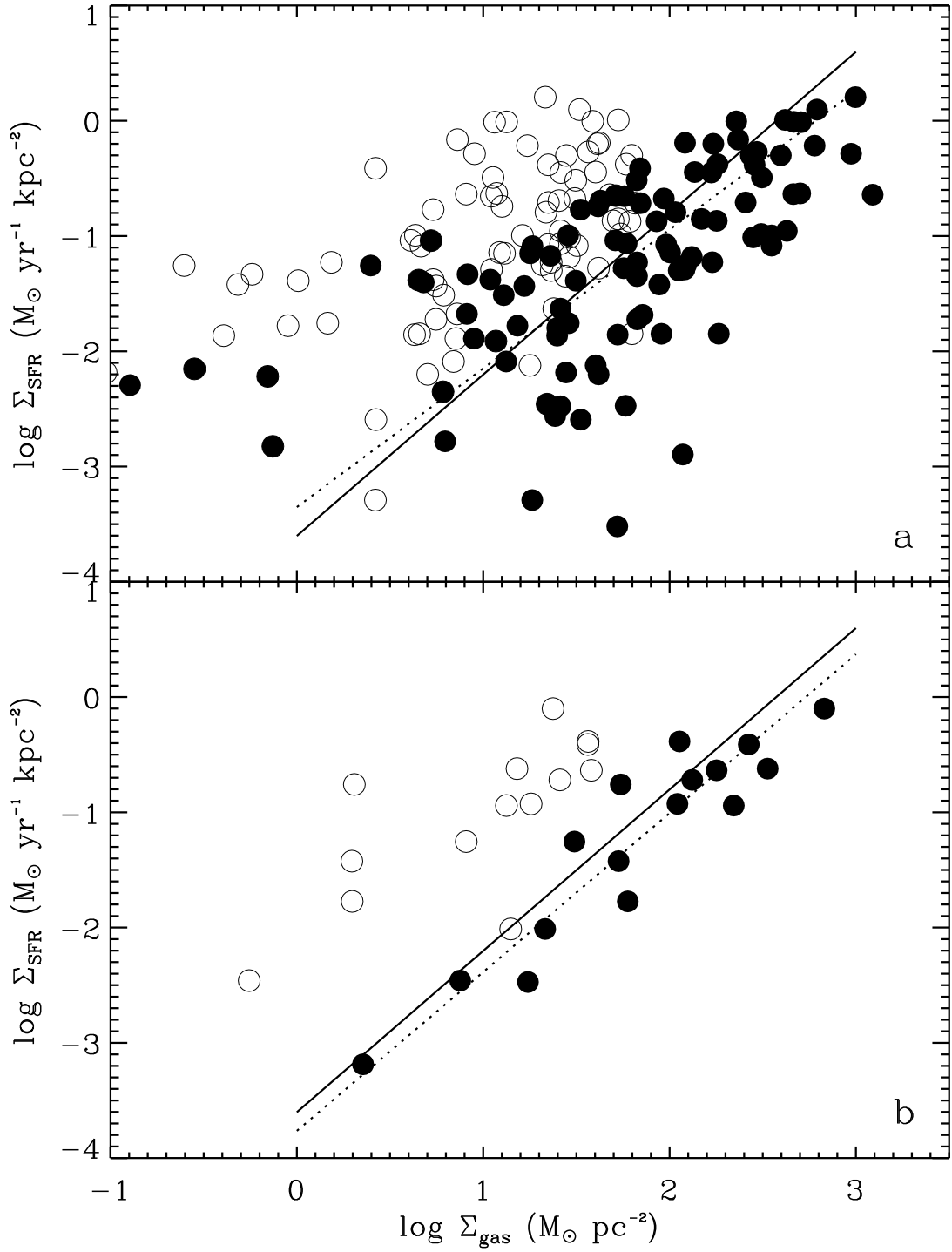
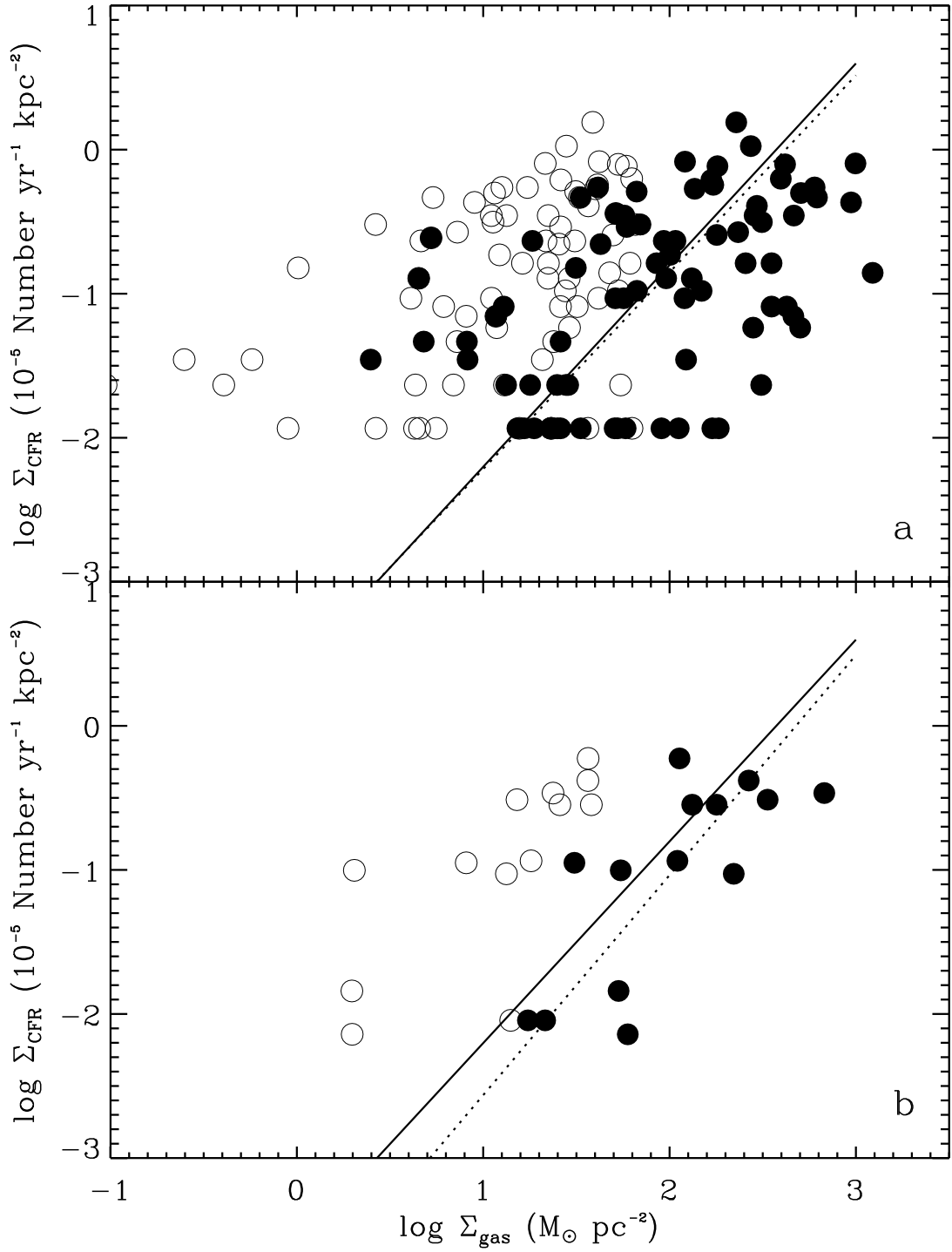
\end{document}